\documentclass[aps,pre,twocolumn,10pt,showkeys,floatfix,preprintnumbers,amsmath,amssymb]{revtex4}
\usepackage{graphicx}% Include figure files
\usepackage{dcolumn}% Align table columns on decimal point
\usepackage{bm}% bold math
\usepackage{amsmath}
\usepackage{enumerate}

\newcommand {\mj} {j}
\begin{document}

\title{Characterization of Fluctuations of Impedance
and Scattering Matrices in Wave Chaotic Scattering}

\author{Xing Zheng}
\author{Sameer Hemmady}
\altaffiliation[Also at ]{Department of Electrical and Computer
Engineering.}
\author{Thomas M. Antonsen Jr. }
\altaffiliation[Also at ] {Department of Electrical and Computer
Engineering.}
\author{Steven M. Anlage}
\author{Edward Ott}
\altaffiliation[Also at ] {Department of Electrical and Computer
Engineering.}\affiliation{Department of Physics\\ and Institute
for Research in Electronics and Applied Physics,\\ University of
Maryland, College Park, MD, 20742}
\date{\today}

\begin{abstract}
In wave chaotic scattering, statistical fluctuations of the
scattering matrix $S$ and the impedance matrix $Z$ depend both on
universal properties and on nonuniversal details of how the
scatterer is coupled to external channels. This paper considers
the impedance and  scattering variance ratios, $VR_z$ and $VR_s$,
where $VR_z=Var[Z_{ij}]/\{Var[Z_{ii}]Var[Z_{jj}] \}^{1/2}$,
$VR_s=Var[S_{ij}]/\{Var[S_{ii}]Var[S_{jj}] \}^{1/2}$, and $Var[.]$
denotes variance. $VR_z$ is shown to be a universal function of
distributed losses within the scatterer. That is, $VR_z$ is
independent of nonuniversal coupling details. This contrasts with
$VR_s$ for  which universality applies only in the large loss
limit. Explicit results are given for $VR_z$ for time reversal
symmetric and broken time reversal symmetric systems. Experimental
tests of the theory are presented using data taken from scattering
measurements on a chaotic microwave cavity.

\end{abstract}

\maketitle                 % Produces the title.

%\newpage

\section{\label{sec:level1}Introduction} % Produces section heading.

The general problem of externally generated time harmonic waves
linearly interacting with a structure of limited spatial extent is
basic in many fields of science \cite{mello04, stockmann99,
haake91}. In recent years much work has been done elucidating the
consequences for the scattering of waves in cases in which, in the
geometric optics approximation, the ray orbits within the
structure are chaotic. Examples  include  optical \cite{newton66},
acoustic \cite{pagneux01}, microwave \cite{doron90, stockmann04,
brouwer97, hemmady04}  and electronic cavities \cite{alhassid00,
altshuler91}. In the case of  complex or irregularly shaped
enclosures  that are large compared with a wavelength, small
changes in the frequency and the configuration give rise to large
changes in the scattering characteristics. This feature  motivates
treatments that are statistical in nature. In this regard random
matrix theory \cite{wigner58, mehta91} has proven useful in
predicting universal aspect of chaotic wave scattering problems in
the cases of both time reversal symmetric systems (corresponding
to matrix statistics of the Gaussian Orthogonal Ensemble, GOE) and
time reversal symmetry broken systems (corresponding to matrix
statistics of the Gaussian Unitary Ensemble, GUE).

Scattering problems can be characterized by the scattering matrix $S$ which relates outgoing scattered wave
amplitudes $b$ to incoming waves $a$, via $b=Sa$. An alternative formulation is in terms of the impedance matrix
$Z$. To illustrate the impedance description, consider an electromagnetic  wave scattering problem in which $N$
transmission lines labeled $i=1,2,\cdots ,N$ of characteristic impedance $Z_{0i}$ are connected to a cavity. Let
$V_i$ and $I_i$ represent the voltage and current on transmission line $i$ as measured at a suitable reference
plane. Then the incident wave $a_i$ and the reflected wave $b_i$ may be expressed as
$a_i=(V_i+Z_{0i}I_i)/Z_{0i}^{1/2}$, $b_i=(V_i-Z_{0i}I_i)/Z_{0i}^{1/2}$. The impedance matrix $Z$ relates the
vector voltage to the vector current, via $V=ZI$, and $Z$ and $S$ are related by
$Z=Z_0^{1/2}(1-S)^{-1}(1+S)Z_0^{1/2}$, where $1$ is the $N$-dimensional identity matrix,  and $Z_0=diag(Z_{01},
Z_{02}, \cdots, Z_{0N})$. The impedance formulation is identical to the   so called ``R-matrix", a formulation
introduced by Wigner and Eisenbud in nuclear-reaction theory in 1947, and further developed in Refs.
\cite{verbaarschot85, lewenkopf91, fyodorov97, beck03}.

\begin{figure}
\includegraphics[scale=0.3, angle=270]{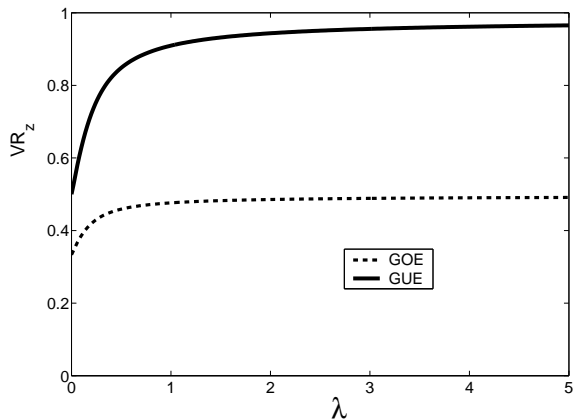}
\caption{$VR_z$ versus the loss parameter $\lambda$, as specified in Eq.~(\ref{eq:VRz_md}) and
Eq.~(\ref{eq:VRz_md2}).} \label{fig:VRz}
\end{figure}

Statistical variations of the elements of $Z$ and $S$ due to small random
variations in the scattering are of great interest. These statistics
have two fundamental influences, (i) universal aspects described by
random matrix theory, and (ii) nonuniversal aspects dependent upon the
details of the coupling of input channels (e.g., transmission lines) to
the scatterer. Our main result concerns the quantity,
\begin{equation}
VR_z=\frac{Var[Z_{ij}]}{\sqrt{Var[Z_{ii}]Var[Z_{jj}]}}, \qquad i\neq j,
\label{eq:VRz}
\end{equation}
where $Var[A]$, the variance of the complex scalar $A$, is
defined as the sum of $Var[ReA]$ and $Var[ImA]$.
Our result is of the  form
\begin{equation}
VR_z=
  \begin{cases}
    F_1(\lambda) & \qquad \text{for GOE,} \\
    F_2(\lambda) & \qquad \text{for GUE,}
  \end{cases}
\label{eq:VRz_general}
\end{equation}
where $\lambda$ is a parameter characterizing the losses within the
scatterer. For example, in the case of an electromagnetic cavity,
$\lambda=\omega/(2Q\Delta \omega)$, where $\omega$ is the frequency of the
incoming signal,
$\Delta\omega$ is the average spacing between cavity resonant frequencies
near
$\omega$, and $Q$ is the quality factor of the cavity
($Q=\infty$  if
there are no internal losses). The remarkable aspect of
(\ref{eq:VRz_general}) is
that
$F_{1,2}(\lambda)$ depends only on  the loss parameter and not on the
nonuniversal properties of the coupling to the cavity. Thus $VR_z$ is a
universal function of the loss $\lambda$. The results for $F_1$ and $F_2$
(to be derived subsequently) are shown in Fig.~\ref{fig:VRz}.
 For
$\lambda \gg 1$,
\begin{equation}
VR_z=
  \begin{cases}
    1/2 & \qquad \text{for GOE, }  \\
    1 & \qquad \text{for GUE. }
  \end{cases}
\label{eq:VRz_highloss}
\end{equation}
For $\lambda \ll 1$,
\begin{equation}
VR_z=
  \begin{cases}
    1/3 & \qquad \text{for GOE, }  \\
    1/2 & \qquad \text{for GUE. }
  \end{cases}
\label{eq:VRz_lowloss}
\end{equation}

A ratio similar to (\ref{eq:VRz}) can also be considered for the
scattering matrix $S$,
\begin{equation}
VR_s \equiv \frac{Var[S_{ij}]}{\sqrt{Var[S_{ii}]
Var[S_{jj}]}}, \qquad i\neq j.
\label{eq:VRs}
\end{equation}
In contrast with (\ref{eq:VRz_general}), $VR_s$ in general depends on {\it
both} the coupling to the cavity and on the loss parameter $\lambda$.
However, in the special case of high loss, $\lambda \gg 1$, $VR_s$ becomes
universal,
\begin{equation}
VR_s=
  \begin{cases}
    1/2 & \qquad \text{for GOE, }  \\
    1 & \qquad \text{for GUE.}
  \end{cases}
\label{eq:VRs_highloss}
\end{equation}
That is, $VR_s=VR_z$ for $\lambda \gg 1$. Based on their
electromagnetics experiments,  Fiachetti and Michelsen
\cite{fiachetti03} have recently conjectured the universality of
(\ref{eq:VRs_highloss}) in the GOE case. More generally,
(\ref{eq:VRs_highloss}) follows from classic results of Hauser and
Feshbach describing fluctuations in the cross section of inelastic
neutron scattering \cite{hauser52}, and this result has been
obtained by Friedman and Mello \cite{mello85_2} using the concept
of maximization of information entropy, and by Agassi {\it et al.}
\cite{agassi75} using a random-matrix model. The important point
is that a universal result for $VR_s$ [i.e.,
Eq.~(\ref{eq:VRs_highloss})] applies only for $\lambda \gg 1$,
while the universal result for $VR_z$, Eq.~(\ref{eq:VRz_general})
and Fig.~\ref{fig:VRz}, is for  arbitrary $\lambda$.

In what follows we derive these results and test them by
comparison with data obtained from scattering measurements on a
chaotic microwave cavity. Section II derives the results for
impedance variance ration, $VR_z$. Section III considers the
scattering variance ratio, $VR_s$. Section IV presents our
experimental tests of the theory.

\section{\label{sec:level2}Impedance Variance Ratio}

 In this section we will obtain the universal functions $F_{1,2}(\lambda)$ in Eq.~(\ref{eq:VRz_general}). We adopt a
formulation (e.g., see Refs. \cite{rcmpaper, rcmpaper2}) that incorporates the nonuniversal effects of the
specific coupling geometry of input-output channels to the scatterer, combined with the random matrix theory for
the universal aspects of the chaotic wave behavior within the scatterer. (In what follows, we use terminology
appropriate to microwave experiments.) Beginning with the case of zero loss ($\lambda=0$), we have that, in the
GOE case, the impedance matrix $Z$ is described by the following statistical model \cite{rcmpaper},
\begin{equation}
Z=-\frac{\mj}{\pi}\sum_n \Delta(k_n) \frac{R_r^{1/2}(k_n) w_n w_n^T R_r^{1/2}(k_n)}{k^2-k_n^2}.
\label{eq:zcav_paper2}
\end{equation}
Here $k$ is the wave number corresponding to the incoming frequency, $w_n$ is a vector whose elements are real,
independent, zero mean, Gaussian random variables of unit variance. The eigenvalues $k_n^2$ are randomly chosen
with statistics appropriate to GOE. This is done by generating ensembles of eigenvalues of random matrices and
then scaling them such that the mean spacing between adjacent eigenvalues near $k_n^2$ is $\Delta(k_n)$. Here
$\Delta(k)$ is given by the Weyl formula for the mean eigenvalue spacing. For example, for the case of a 2D cavity
$\Delta(k)$ is independent of $k$, $\Delta=4\pi/A$ where $A$ is the area of the cavity. The system-dependent part
of the coupling is characterized by the corresponding radiation impedance matrix $Z_r=R_r+jX_r$. Note also that,
since $Z_r$ depends only on local geometry near the entrances to the scattering region, its frequency dependence
is much slower than that of $Z$ which depends on the geometry of the scattering region assumed to be characterized
by much longer length scales. The radiation impedance is the impedance seen at the reference plane of the input
channel when waves that leave the channel propagate outward and are not reflected back to their input by the
distant walls of the cavity. Thus the radiation impedance is the impedance seen when the distant walls are removed
to infinity, or (as can be done in an experimental measurement \cite{hemmady04}) when the distant walls are lined
with absorber. Hence the radiation impedance depends only on the local structure in the vicinity of the port
coupling to the cavity and not on the shape or chaotic properties of the cavity. In the case of ports that are far
apart, e.g., of the order of the cavity size, the off-diagonal elements of $Z_r$ are small and will be neglected.
Thus we will take $Z_r$ to be a diagonal matrix with elements $Z_{ri}=R_{ri}+jX_{ri}$. In the GUE case, the
elements of $w_n$ in (\ref{eq:zcav_paper2}) are complex with real and imaginary parts each individually having
Gaussian statistics, $w_n^T$ in (\ref{eq:zcav_paper2}) is replaced by $w_n^{\dag}$ (where $\dag$ denotes the
conjugate transpose), and the statistics of $k_n^2$ are now those appropriate to GUE.

Equation (\ref{eq:zcav_paper2}) is derived in \cite{rcmpaper}. It
results from a formal series expansion of the solution for $Z$ in
terms of eigenfunctions of the closed cavity. The local structure of
the eigenfunctions is then assumed to satisfy the so-called "random
plane wave hypothesis"; namely, that in the semi-classical limit
($k^{-1}$ is large compared to the typical dimensions of the
structure), if ray trajectories in the closed cavity are chaotic,
then eigenfunction $n$ is statistically similar to a random
isotropic superposition of plane waves of wave number $\vec{k}$ with
magnitude$|\vec{k}|=k_n$. This gives Eq.~(\ref{eq:zcav_paper2})
\cite{rcmpaper}. We then further supplement
Eq.~(\ref{eq:zcav_paper2}) with the known distribution function of
$k_n^2$ determined from random matrix theory and the local mean
spacing $\Delta k_n^2$ between adjacent eigenvalues.

The universality of the result for $VR_z$ (Fig.~1) can be shown from (\ref{eq:zcav_paper2}) as follows. The
impedance matrix $Z$ defined in (\ref{eq:zcav_paper2}) will have a  mean and a fluctuating component. As shown in
refs \cite{rcmpaper, rcmpaper2}, the value of the mean is determined by all terms in the sum and thus depends on the
slow $k$-dependence of $R_r(k)$ and $\Delta(k)$, while the fluctuating component of $Z$ is determined by terms in
the sum in Eq.~(\ref{eq:zcav_paper2}) for which $k^2\simeq k_n^2$. Thus if $R_r(k_n)$ and $\Delta(k_n)$ are
approximately constant over a range of $k_n^2$ values corresponding to  many resonances, then the fluctuating part
of $Z$ will be universal after appropriate normalization. If we define $\tilde{Z}$ to be the fluctuating part of
$Z$, we have
\[
R_r^{-1/2}(k)\tilde{Z}R_r^{-1/2}(k)=-\frac{j}{\pi}\sum_n \frac{w_nw_n^{\dagger}}{s_n}\equiv \zeta, \qquad \quad
(7')
\]  where $s_n$ are the eigenvalues of a GOE random matrix, normalized to have unit mean spacing between adjacent
eigenvalues near $s_n\simeq 0$. Thus $\zeta$ is a universal normalized impedance matrix. If $Z_r$ is diagonal, we
have $\zeta_{ij}=R_{ri}^{-1/2}R_{rj}^{-1/2}Z_{ij}$. Hence the ratios $VR_z$ defined by (\ref{eq:VRz}) is the same
for $Z$ and $\zeta$, and is, therefore, universal. (We emphasize, however, that this conclusion relies on $Z_r$ being
diagonal.)

The effects of distributed loss, such as losses due to conducting walls or a lossy dielectric that fills the
cavity, can be simply incorporated  in Eq.~(\ref{eq:zcav_paper2}).  Since modal fluctuations in  losses are small
when the modes are chaotic and the wavelength is short, we can construct a complex cavity impedance accounting for
distributed loss by simply replacing $k^2$ in Eq.~(\ref{eq:zcav_paper2}) by $k^2(1-jQ^{-1})$, where $Q$ is the
cavity quality factor. (In terms of the normalized impedance matrix $\zeta$ defined in ($7'$) we replace the
denominator $s_n$ by $s_n-jk^2/[Q\Delta(k)]$.) For example,
\begin{equation}
\begin{aligned}
&Z_{ii}=-\frac{j}{\pi}\sum_{n=1}\frac{R_{ri}\Delta_n
w_{in}^2}{k^2(1-jQ^{-1})-k_n^2}\equiv R_{ii}+jX_{ii}
 \\
&=\frac{1}{\pi}[\sum_{n=1}\frac{R_{ri}\Delta_n w_{in}^2
k^2/Q}{(k^2-k_n^2)^2+(k^2/Q)^2}+j\sum_{n=1}\frac{R_{ri}\Delta_n w_{in}^2
(k_n^2-k^2)}{(k^2-k_n^2)^2+(k^2/Q)^2}].
\end{aligned}
\label{eq:z11_goe}
\end{equation}
(Henceforth, we employ the notation $\Delta_n\equiv \Delta(k_n)$.) Calculation of the moments of the impedance is
facilitated by the fact that the eigenvalues and eigenfunctions in the chaotic cavities are statistically
independent. For example, the expected value of $X_{ii}$ is,
\begin{equation}
\begin{aligned}
E[X_{ii}]&=\lim_{M\rightarrow \infty}\frac{1}{\pi}\sum_{n=1}^M\int
dw_{in} f(w_{in}) w_{in}^2
\int dk_1^2  \cdots dk_M^2\\
& P_J(k_1^2, \cdots , k_M^2)
\frac{R_{ri}\Delta_n (k_{n}^2-k^2)}{(k^2-k_{n}^2)^2+(k^2/Q)^2},
\end{aligned}
\label{eq:e1x11_joint}
\end{equation}
where $f(w_{in})$ is the probability distribution function (pdf) of
$w_{in}$ and $P_J$ is the joint pdf of the eigenvalues.
Integrating over all $k_j$, $j\neq n$, we express $E[X_{ii}]$ as an
integral over the pdf of $k_n^2$,
$P_1(k_{n}^2)=1/(\Delta_n M)$,
we  consider the $M\rightarrow \infty$ limit and use $\langle w_n^2
\rangle=1$
for the Gaussian random variable $w_n$.
\begin{equation} \begin{aligned}
E[X_{ii}]
=\int
dk_{n}^2 \frac{R_{ri}(k_n)(k_{n}^2-k^2)/\pi
}{(k^2-k_{n}^2)^2+(k^2/Q)^2}=X_{ri}(k).
\end{aligned}
\label{eq:e1x11}
\end{equation}
The second equality in (\ref{eq:e1x11}) relating $E[X_{ii}]$ to
the radiation reactance requires  $Q\gg 1$ and is analogous to the
Kramers-Kronig relation.

The second moment of $X_{ii}$ can be determined in a similar way by integrating over all $j$ except $j=t,s$ and
using the joint distribution function $P_2(k_t^2, k_s^2)=[1-g(|k_t^2-k_s^2|/\Delta)]/(M\Delta)^2$, where
$g(|k_t^2-k_s^2|/\Delta)$ is  known from Random Matrix theory \cite{mehta91}. Using the fact that $g$ goes to zero
at large argument and assuming that the radiation resistance $R_{ri}(k_n)$ and the average spacing $\Delta_n$ vary
slowly over the damping width $k^2/Q$, we obtain
\begin{equation}
\begin{aligned}
&E[X_{ii}^2]=
\frac{3}{2\pi}(R_{ri}^2\frac{\Delta}{k^2/Q})+\{E[X_{ii}]\}^2
+\frac{R_{ri}^2}{\pi^2}\int dk_t^2 dk_s^2 \\
& \frac{\langle w^2 \rangle ^2
g(|k_t^2-k_s^2|/\Delta)(k_t^2-k^2)(k_s^2-k^2)}{[(k_t^2-k^2)^2+(k^2/Q)^2]
[(k_s^2-k^2)^2+(k^2/Q)^2]},
\end{aligned} \label{eq:e2x11}
\end{equation}
Combining Eq.~(\ref{eq:e2x11})
and Eq.~(\ref{eq:e1x11}), we obtain
\begin{equation}
Var[X_{ii}]=\frac{R_{ri}^2}{\lambda}[\frac{3}{2\pi}-\frac{1}{\pi}
\int_0^{\infty}
d x g(x) \frac{4}{4+(x/\lambda)^2}],
\label{eq:var_x11_simp}
\end{equation}
where $\lambda=k^2/(Q\Delta)$. A similar moment evaluation can be carried out for $R_{ii}$, as specified in
Eq.~(\ref{eq:z11_goe}), which  yields  the same expression as Eq.~(\ref{eq:var_x11_simp}) for $Var[R_{ii}]$. For
GOE (the case we are now considering) we have that \cite{mehta91}, $g(s)=f^2(s)-[\int_0^s d(s')f(s')-1/2](d f/d
s)$, where $f(s)=[(\sin \pi s)/(\pi s)]$.

In order to obtain the variance ratio, we also apply the previous
process to the off diagonal term $Z_{ij}$, $i\neq j$,  which, based on
Eq.~(\ref{eq:zcav_paper2}), is given by
\begin{equation} \begin{aligned}
Z_{ij}&=\frac{1}{\pi}[\sum_n
\frac{\sqrt{R_{ri}R_{rj}}\Delta_n
w_{in}w_{jn}k^2/Q}{(k^2/Q)^2+(k^2-k_n^2)^2} \\
& + j\sum_n
\frac{\sqrt{R_{ri}R_{rj}}\Delta
w_{in}w_{jn}k^2/Q}{(k^2/Q)^2+(k^2-k_n^2)^2}]. \end{aligned}
\label{eq:z12} \end{equation} Since $w_{in}$ and $w_{jn}$ are independent,
the first moments of $X_{ij}$ and $R_{ij}$ are both zero, and the variance
is equal to the second moment, \begin{equation} \begin{aligned}
Var[X_{ij}]&=\lim_{M\rightarrow
\infty}\frac{M}{\pi^2}\int dk_{n}^2
\frac{R_{ri}R_{rj}\Delta_n^2\langle w_{in}^2\rangle \langle
w_{jn}^2\rangle}
{[(k_{n}^2-k^2)^2+(k^2/Q)]^2}P_1(k_{n}^2) \\
&=\frac{R_{ri}R_{rj}}{\lambda}\frac{1}{2\pi}, \end{aligned}
\label{eq:var_x12} \end{equation} The same result is obtained for
$Var[R_{ij}]$. Combining Eq.~(\ref{eq:var_x12}) with
Eq.~(\ref{eq:var_x11_simp}), we have Eq.~(\ref{eq:VRz_general}) with
\begin{equation} \begin{aligned} VR_z =F_1(\lambda)=[3-2\int_0^{\infty} d x
g(x)\frac{4}{4+(x/\lambda)^2}]^{-1}. \end{aligned} \label{eq:VRz_md}
\end{equation}

A similar calculation in the GUE case is facilitated by the simpler form
of the function $g(x)$ which is now given by $g(x)=\sin^2(\pi x)/(\pi
x)^2$. We obtain
\begin{equation}
\begin{aligned}
VR_z&=F_2(\lambda)=[2- 2\int_0^{\infty}dx
(\frac{\sin \pi x}{\pi x})^2 \frac{4}{4+(x/\lambda)^2}]^{-1} \\
&=[1+\frac{1-e^{-4 \pi \lambda}}{4 \pi \lambda}]^{-1}.
\end{aligned}
\label{eq:VRz_md2}
\end{equation}
%Performing the integral in (\ref{eq:VRz_md2}), we get
%\begin{equation}
%\label{eq:VRz_md2_inte}
%\end{equation}

We note that subsequent to a previous announcement of our work
\cite{arxiv_this}, the two-frequency correlation functions for the
elements of the impedance and the scattering matrix have recently
been calculated by Savin, Fyodorov and Sommers \cite{savin05}, and
are consistent with the preceding in the limit of zero frequency
separation.

\section{Scattering Variance Ratio}

We now consider the
scattering matrix in the high loss limit, $\lambda \gg 1$. For simplicity, we
consider the case of two channels connecting to the scatterer, $N=2$, and $Z$ and
$S$ are $2 \times 2$ matrices. We note that a chaotic
scattering process can be divided into a direct process and a delayed
process, which leads to a separation of the mean part (equal to $Z_r$)
and the fluctuating part $\delta Z$  of the cavity impedance,
$Z=Z_r+\delta Z$.
The fluctuating part $\delta Z$ decreases as loss
increases. Thus in the high loss limit, $\delta Z \ll Z_r$, which
implies $Z_{12}, Z_{21} \ll Z_{11},Z_{22}$. (Recall,  the mean parts of
the off
diagonal components are zero.) We may now form $S$ using
$S=Z_0^{-1/2}(Z-Z_0)(Z+Z_0)^{-1}Z_0^{1/2}$. Since the off diagonal terms of
$Z$ are small, the diagonal components of $S$ are dominated by the diagonal
components of $Z$. We then find for $S_{11}$,
\begin{equation}
\begin{aligned} S_{11}&\cong \frac{Z_{11}-Z_{01}}{Z_{11}+Z_{01}}=
\frac{(Z_{r1}-Z_{01})+\delta Z_{11}}{(Z_{r1}+Z_{01})+\delta Z_{11}}\\
&\cong S_{r1}+[\frac{2Z_{01}}{(Z_{r1}+Z_{01})^2}]\delta Z_{11},
\end{aligned} \label{eq:s11dz} \end{equation} where
$S_{r1}=(Z_{r1}-Z_{01})/(Z_{r1}+Z_{01})$, and $Z_{01}$ is the
characteristics impedance of  channel 1. Thus, we obtain
\begin{equation}
Var[S_{11}]=|\frac{2Z_{01}}{(Z_{r1}+Z_{01})^2}|^2 Var[Z_{11}].
\label{eq:var_s11} \end{equation} In addition, we can express
$S_{12}$ in the high damping limit as \begin{equation}
\begin{aligned}
S_{12}&=\frac{2Z_{12}\sqrt{Z_{01}Z_{02}}}{(Z_{11}+Z_{01})(Z_{22}+Z_{02})}
 \simeq
\frac{2Z_{12}\sqrt{Z_{01}Z_{02}}}{(Z_{r1}+Z_{01})(Z_{r2}+Z_{02})},
\label{eq:s12dz} \end{aligned} \end{equation} which  leads to
\begin{equation}
Var[S_{12}]=|\frac{2\sqrt{Z_{01}Z_{02}}}{(Z_{r1}+Z_{01})(Z_{r2}+Z_{02})}|^2
Var[Z_{12}], \label{eq:var_s12} \end{equation}
and similarly for $Var[S_{21}]$. Combining  Eq.~(\ref{eq:var_s11}) and
Eq.~(\ref{eq:var_s12}), we
recover Eq.~(\ref{eq:VRs_highloss}) and we note that this result is
independent of the coupling (i.e., independent of $Z_r$).

\begin{figure}
\includegraphics[scale=0.25]{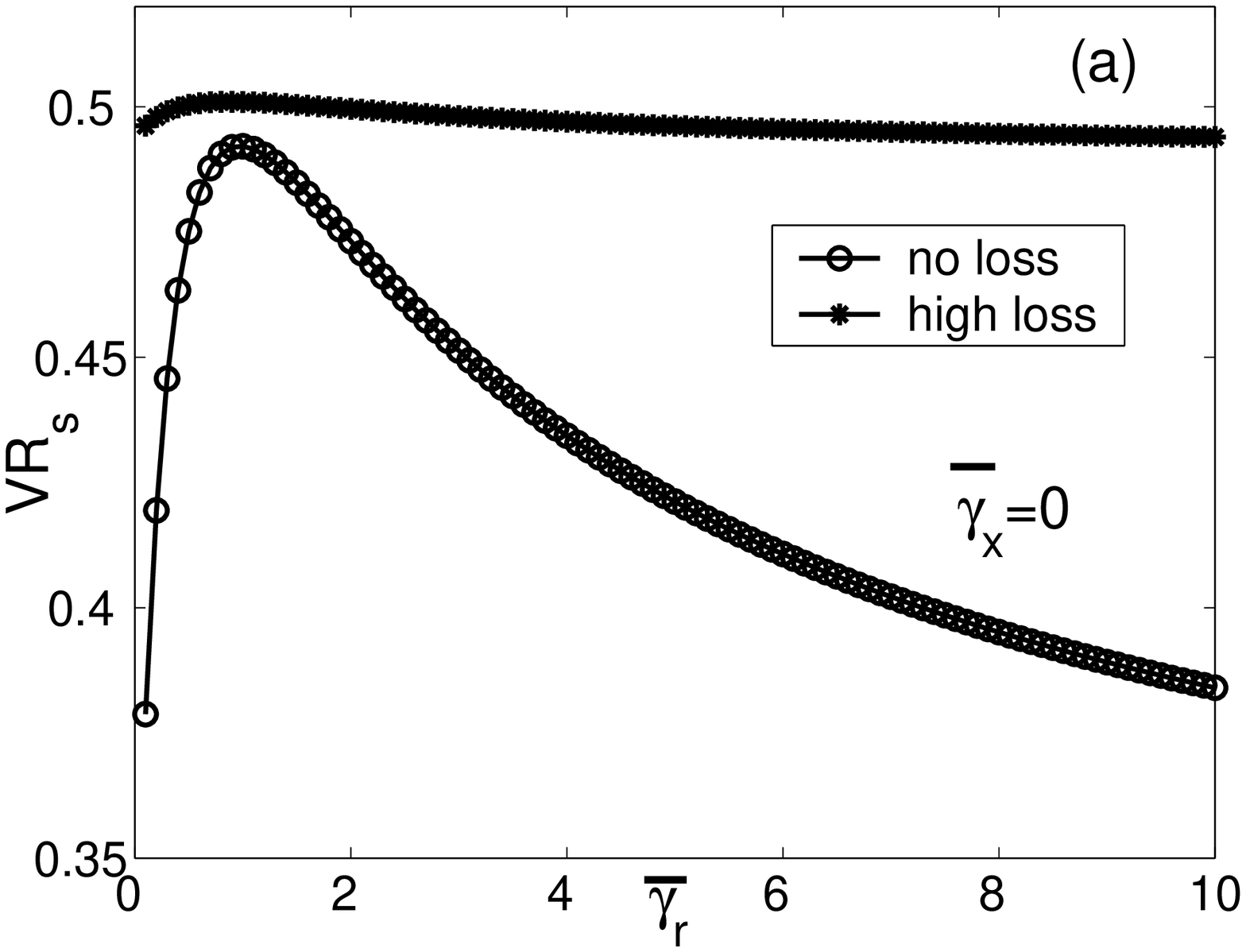}
\includegraphics[scale=0.25]{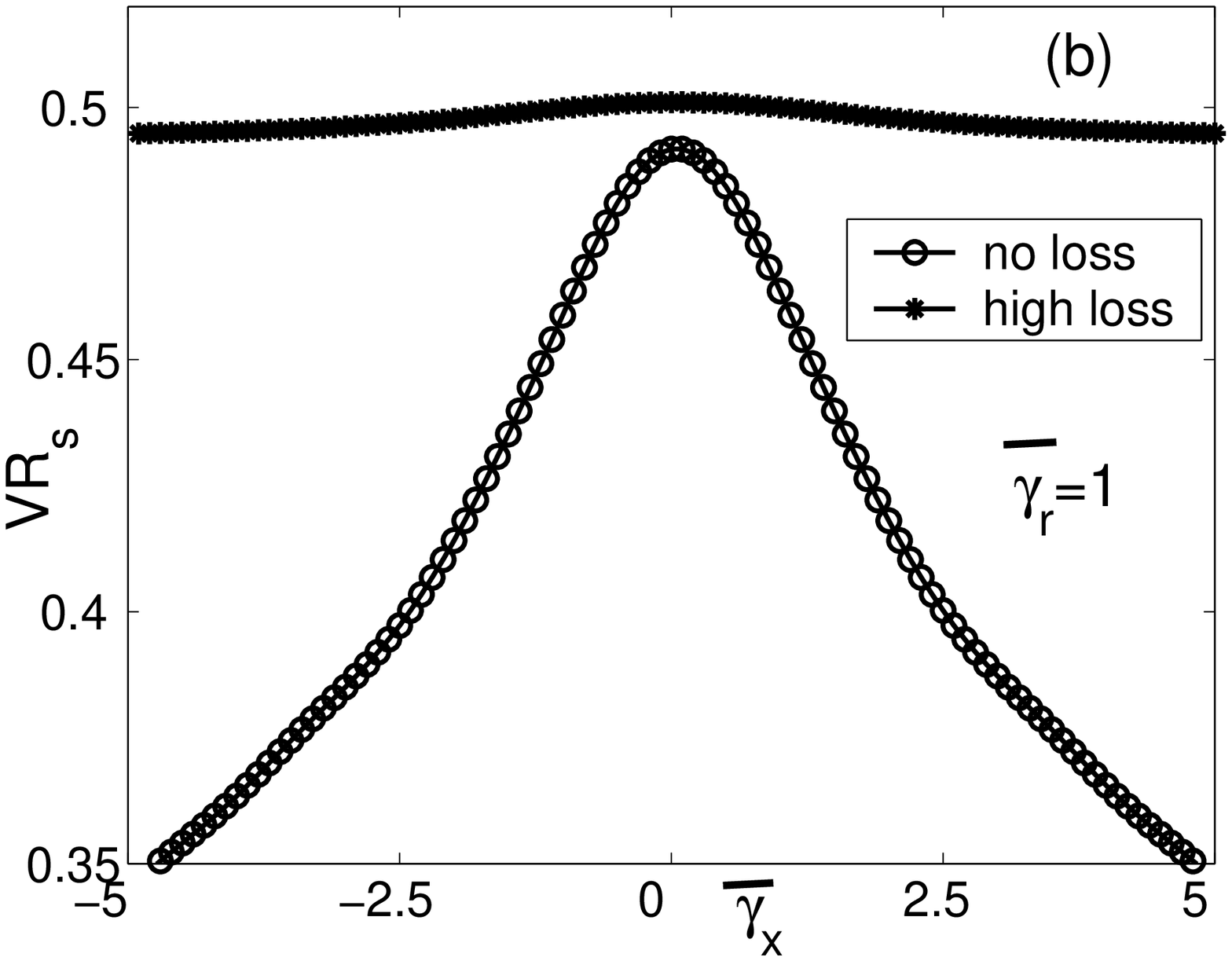}
\caption{(a) $VR_s$ versus $\bar \gamma_r$ for $\bar \gamma_x=0$ in the lossless case $\lambda=0$ and in a high
loss case $\lambda=5$. (b)$VR_s$ versus $\bar \gamma_x$ for $\bar \gamma_r=1$.} \label{fig:nonpcVRs}
\end{figure}

To illustrate the influence  of coupling on  $VR_s$ at finite loss
parameter $\lambda$, we consider the impedance matrix in the GOE
case using the model normalized impedance $\zeta$ used in
Ref.~\cite{rcmpaper}, $Z=R_r^{1/2}\zeta R_r^{1/2}+jX_r$, where
$\zeta$ is given by
$\zeta_{ij}=-(j/\pi)\sum_{n=1}^{M}(w_{in}w_{jn})/(\tilde
k^2-\tilde k_n^2 -j\lambda)$, $\tilde k_n^2=k^2/\Delta$, and
$\tilde k^2$ is set to be $M/2$, such that mean of $\zeta$ is
zero. Realizations of $\zeta$ are produced numerically by
generating Gaussian random  variables $w_{in}$ and spectra $\tilde
k_n^2$ from the eigenvalues of random matrices. We  express a
model scattering matrix $S$ as
\begin{equation}
S=(\gamma_r^{1/2}\zeta\gamma_r^{1/2}+j\gamma_x+1)^{-1}
(\gamma_r^{1/2}\zeta\gamma_r^{1/2}+j\gamma_x-1),
\label{eq:modelS}
\end{equation}
where $\gamma_r=Z_0^{-1}R_r$ and $\gamma_x=Z_0^{-1}X_r$, When $\gamma_r$ is the identity matrix and $\gamma_x$ is
zero, we reach the so-called perfect coupling condition, which means that the scattering is determined  by the
delayed process and the direct process is absent. We now consider an example in which the two-port couplings are
the same so that $\gamma_{r,x}=diag(\bar\gamma_{r,x}, \bar\gamma_{r,x})$, where $\bar\gamma_{r,x}$ is a scalar.
Figures \ref{fig:nonpcVRs}(a) and (b) show results for the variation of $VR_s$ with the coupling parameters
$\bar\gamma_r$ and $\bar\gamma_x$, for a high loss case ($\lambda=5$) and for the lossless case ($\lambda=0$). In
Fig.~\ref{fig:nonpcVRs}(a), we fix $\bar \gamma_x$ to be zero, and vary $\bar \gamma_r$, while in
Fig.~\ref{fig:nonpcVRs}(b), $\bar \gamma_r$ is fixed to be 1 and $\bar \gamma_x$ is varied.  Compared to the high
damping case,  $VR_s$ in the lossless case has a much larger deviation from the constant 1/2. Note that $VR_s$ is
$1/2$ in the  perfect-coupling case (i.e., $\bar \gamma_r=1$, $\bar \gamma_x=0$), no matter whether the cavity is
highly lossy or lossless. This is related to the concept of ``weak localization" reviewed in \cite{lee85}.

In the case of an $N$-port we can think of the above two-port consideration of $VR_s$  as  applying to the
$N$-port converted to a two port by opening channels $3,4,\cdots, N$; i.e., the incoming waves $a_3, a_4, \cdots
a_N$ are identically zero (for a microwave cavity with transmission line inputs, this corresponds to terminating
transmission lines $3,4,\cdots, N$ with their characteristic impedances, $Z_{03}, Z_{04}, \cdots, Z_{0N}$). Thus
ports $3,4,\cdots, N$  effectively add to the loss due to the energy flux leaving through them. If the ports
$3,4,\cdots, N$ are assumed to act like distributed loss, they can be taken into account by increasing the loss
parameter $\lambda$. [This increased loss enhances the validity of Eq.~(\ref{eq:VRs_highloss}).]

\section{\label{sec:level4}Experimental Tests}
We provide experimental results testing the theoretical predictions for the statistical fluctuations in the
variance of the $S$ and $Z$ elements, in the limit of large damping. The experiments are done in an air-filled,
quarter bow-tie shaped cavity which acts as a two-dimensional resonator below 19.05 GHz
(Fig.~3(a))\cite{gokirmak98}. This cavity has previously
 been used for the successful study of the eigenvalue spacing statistics \cite{so94} ,
eigenfunction statistics \cite{wu98, chung00}, and for studying the universal fluctuations in the impedance
\cite{hemmady04} and scattering matrix \cite{hemmady05} for a wave chaotic system. The cavity is driven by two
ports; each of which consists of the center conductor (diameter 2a=1.27mm) of a coaxial cable that extends from
the top lid of the cavity and makes contact with the bottom plate of the cavity (Fig.~3(b)). From direct $S_{21}$
measurements, we estimate that the cavity has a typical loaded $Q$ of about 150 between 4-5GHz and about 300
between 11-12GHz. This translates to a damping parameter of $\lambda > 0.5$ for the entire frequency range of this
experiment. Hence we examine experimentally the time-reversal symmetric (GOE) cases for the $Z$ and $S$-variance
ratios in the high damping limit.
\begin{figure}
\includegraphics[scale=0.5, angle=270]{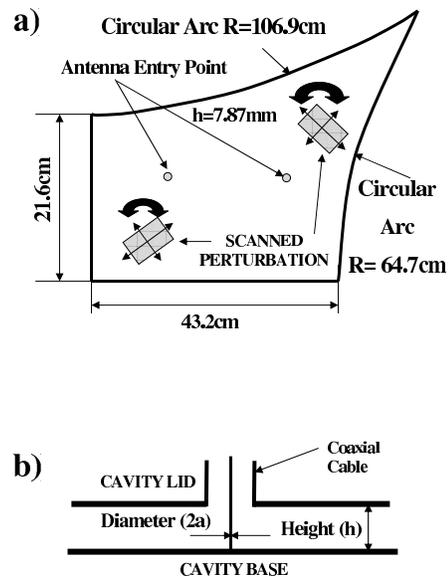}
\caption{(a) The physical dimensions of quarter bow-tie chaotic
microwave resonator are shown  along with the position of the two
coupling ports. Two metallic perturbers are shown in gray. (b) The
details of the coupling ports (antennas) and cavity height $h$ are
shown in cross section. } \label{fig:samy1}
\end{figure}

To perform an approximation to ensemble averaging,  two perturbers
(shown gray in Fig.~3(a)), made up of rectangular ferromagnetic
solids
 wrapped in Al foil (of dimensions 26.7 x 40.6 x 7.87 $mm^3$), are systematically
 scanned and rotated throughout the volume of
the cavity by means of a magnet that is placed outside the cavity.
The ensemble set consists of one-hundred different positions and
orientations of the perturbers within the cavity. The perturbers are
on the scale of half the wavelength or bigger over the entire
frequency range of the experiment. The full two by two $S$ matrix is
measured between 4 and 12 GHz for each position of the perturbers.
Below 4 GHz the mode density is too low to obtain meaningful
ensemble averaging, while
 above 12 GHz the coupling becomes too weak to couple to the modes of
  the cavity, at least for this port geometry.
Once the $S$ matrix has been measured, it is then converted to $Z$ through $Z=Z_0 (I+S)(I-S)^{-1}$.

To eliminate the effect on the average of short ray orbits returning
to the antenna (these lead to rapidly varying-with-frequency
systematic deviations of the average from the ensemble average, as
discussed in \cite{rcmpaper}) we perform frequency averaging over a
sliding window of width 300 MHz.  We denote such sliding averages of
impedance and scattering variance ratios by $\overline{VR}_z$ and
$\overline{VR}_s$. The inset in Fig.~4 shows $\overline{VR}_z$
(solid line) over a frequency range 4 GHz to 12 GHz. Denoting the
average of $\overline{VR}_z$ over the entire range, 4GHz-12GHz, by
$\overline{\overline{VR}}_z$, we obtain
$\overline{\overline{VR}}_z=0.49$, and we find that
$|\overline{VR}_z - \overline{\overline{VR}}_z|\leq 0.02$ over the
entire frequency range. This value of experimentally obtained
$\overline{ \overline{VR}}_z$ is close to the ideal theoretical
value of 1/2 for large damping. Also shown in the inset is the
variance ratio obtained with no frequency averaging (small circles)
These are deviations from the frequency averaged ratio values with a
standard deviation of 0.04. Nevertheless the mean value of the
variance ratio over the entire frequency range is 0.49.

The circles, stars and dashes in Fig. 4 show the variation in
$\ln[Var[Z_{21}]]$, $\ln[\sqrt{Var[Z_{11}]Var[Z_{22}]}]$ and
$\ln[\overline{\overline{VR}}_z\sqrt{Var[Z_{11}]Var[Z_{22}]}]$
respectively, as a function of frequency. The agreement is quite
good (i.e., the dashes overlie the open circles) at all frequencies.
\begin{figure}
\includegraphics[scale=0.50, angle=0]{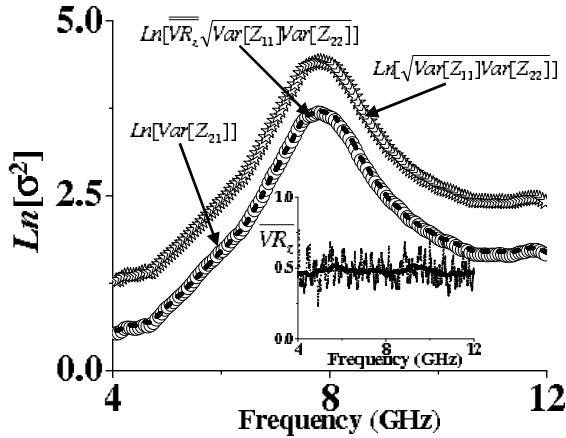}
\caption{$Var[Z_{21}]$(circles),
$\sqrt{Var[Z_{11}]Var[Z_{22}]}$(stars) and
$\overline{\overline{VR}}_z\sqrt{Var[Z_{11}]Var[Z_{22}]}$(dashes)
are plotted  on a natural-logrithmic scale as a function of
frequency from 4 to 12 GHz. Inset shows the ratio $\overline{VR}_z$
versus frequency (solid line). The small circles show $VR_z$ without
any frequency averaging.}\label{fig:samy2}
\end{figure}

 Similarly in Fig.~5, we present data for the scattering variance ratio. Experimentally we obtain
$\overline{\overline{VR}}_s=0.50$ and $|\overline{VR}_s - \overline{\overline{VR}}_s|\leq 0.08$ over the range 4
GHz to 12 GHz. The circles, stars and dashes in Fig.~5 show the variation in $Var[S_{21}]$,
$\sqrt{Var[S_{11}]Var[S_{22}]}$ and $\overline{\overline{VR}}_s \sqrt{Var[S_{11}]Var[S_{22}]}$, respectively, as a
function of frequency. Similar to the impedance data (Fig.~\ref{fig:samy2}), we observe that the data for
$\overline{\overline{VR}}_s \sqrt{Var[S_{11}]Var[S_{22}]}$(dashes) overlie the data for $Var[S_{21}]$(open
circles), again indicating that the experimentally obtained value for
$\overline{\overline{VR}}_s$ shows good
agreement with the asymptotic theoretical values for highly damped time-reversal symmetric systems over a large
frequency range.

\begin{figure}[b]
\centering
\includegraphics[scale=0.35, angle=270]{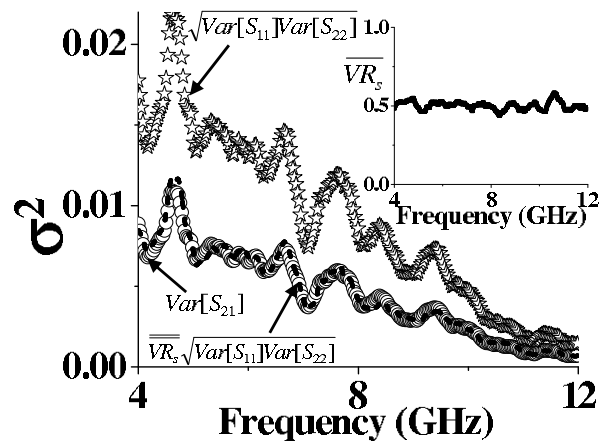}
\caption{$Var[S_{21}]$(circles), $\sqrt{Var[S_{11}]Var[S_{22}]}$(stars) and
$\overline{\overline{VR}}_s\sqrt{Var[S_{11}]Var[S_{22}]}$(dashes) are plotted as a function of frequency from 4 to
12 GHz. Inset shows the ratio $\overline{VR}_s$ versus frequency.}\label{fig:samy3}
\end{figure}

To sum up, we have used  random matrix theory  in conjunction with the
radiation impedance characterizing the
system-dependent coupling details to evaluate the variance ratios
(\ref{eq:VRz}) and (\ref{eq:VRs}). The main result is that the impedance
variance ratio (\ref{eq:VRz}) is a universal function of the loss in the
scatterer.

This work was originally stimulated by discussion with J. P.
Parmantier who pointed out to us the result of
Ref.~\cite{fiachetti03}. We also acknowledge discussions with P.
A. Mello  and D. V. Savin. This work was supported in part by the
DOD MURI for the study of microwave effects under AFOSR Grant
F496200110374, and DURIP grants FA95500410295 and FA95500510240.

%In order to obtain the variance of $R(\sigma)$, we  calculate the
%second moment of

\end{document}